\documentstyle[emulateapj,danonecolfloat,epsfig]{article}

\newcommand\lsim{\mathrel{\rlap{\lower4pt\hbox{\hskip1pt$\sim$}}
        \raise1pt\hbox{$<$}}}
\newcommand\gsim{\mathrel{\rlap{\lower4pt\hbox{\hskip1pt$\sim$}}
        \raise1pt\hbox{$>$}}}

\begin{document}
\twocolumn[


\tighten
\eqsecnum
\received{May 26 2003}
\accepted{December 22 2003}
\journalid{605}{20 April 2004}
\articleid{11}{14}


\title{Beyond Lyman $\alpha$: Constraints and Consistency Tests from the Lyman $\beta$ Forest}

\author{Mark Dijkstra$^1$,Adam Lidz$^2$ and Lam Hui$^{2,3}$}
\affil{$^1$ Department of Astronomy, Columbia University, New York, NY 10027\\
$^2$ Department of Physics, Columbia University, New York, NY 10027\\
$^3$ Theoretical Astrophysics, Fermi National Accelerator Laboratory, Batavia, IL 60510\\
Electronic mail: {\tt mark,lidz,lhui@astro.columbia.edu}}

\begin{abstract}
Absorption between the rest-frame wavelengths of $973 \AA$ and 
$1026 \AA$ in quasar spectra arises from two sources (apart from
occasional metals): one is due to 
Ly$\alpha$ absorption by materials at a low redshift, and the other is from Ly$\beta$ at a higher
redshift. These two sources of absorption are to a good approximation
uncorrelated because of their wide physical separation.
Therefore, the two-point correlation of absorption in this region of quasar spectra
neatly factorizes into two pieces: the Ly$\beta$ correlation
at high $z$, and the Ly$\alpha$ correlation at low $z$. 
The latter can be independently measured from quasar spectra at lower redshifts
using current techniques. A simple division then offers a way to statistically 
separate out the Ly$\beta$ two-point correlation from the Ly$\alpha$ correlation.
Several applications of this technique are discussed.
First, since the Ly$\beta$ absorption cross-section is lower than Ly$\alpha$
by about a factor of $5$, the Ly$\beta$ forest is a better probe of the 
intergalactic medium (IGM) at higher redshifts where Ly$\alpha$ absorption is often saturated.
Second, for the same reason, the Ly$\beta$ forest allows a better measurement
of the equation of state of the IGM at higher overdensities, yielding stronger
constraints on its slope when used in conjunction with the Ly$\alpha$ forest.
Third, models of the Ly$\alpha$ forest based on gravitational instability
make unique predictions for the Ly$\beta$ forest, which can be tested against
observations. We briefly point out that feedback processes which affect higher density regions but leave
low density structure intact may be better constrained by the Ly$\beta$ forest.
Finally, extending our technique to the higher Lyman series is in principle possible,
but becomes increasingly difficult because of diminishing path lengths.
\end{abstract}

\keywords{cosmology: theory -- intergalactic medium -- large scale
structure of universe; quasars -- absorption lines}
]

\section{Introduction}
\label{intro}

\begin{figure*}[t]
\vbox{ \centerline{ \epsfig{file=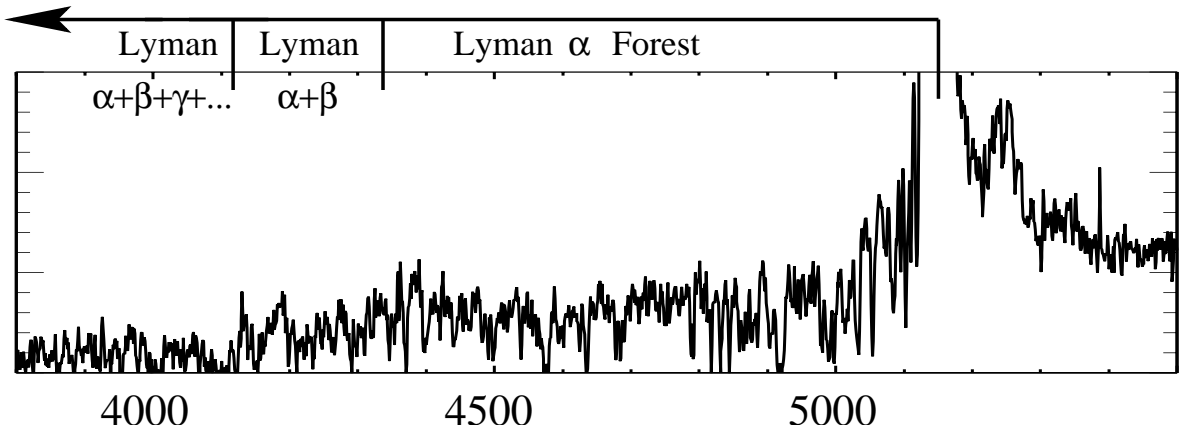,width=18.0truecm}}
\figcaption[]{ \footnotesize Schematic diagram to show the relevant regions discussed in
 this paper.  The region in which only Ly$\alpha$ photons can be absorbed: 
$((1+z_q)\lambda_{\beta}^0,(1+z_q)\lambda_{\alpha}^0)$, where $z_q$ is the 
quasar redshift, is the famous Ly$\alpha$ forest. 
The focus of this paper is on the Lyman $\alpha+\beta$ 
region, $((1+z_q)\lambda_{\gamma}^0,(1+z_q)\lambda_{\beta}^0)$. The spectrum 
is of the quasar Q2139-4434 at $z_q=3.23$, kindly provided by Arlin Crotts. 
Note that the rest-frame Ly$\alpha$, Ly$\beta$ and Ly$\gamma$ 
wavelengths are: $\lambda_{\alpha}^0=1215.67
\AA$,$\lambda_{\beta}^0=1025.72 \AA$ and $\lambda_{\gamma}^0=972.54 \AA$. The Lyman $\alpha$ emission line
at $\lambda=5144 \AA$ is not fully shown.
\label{sch}}}
\end{figure*}

A typical high redshift quasar spectrum is shown in Fig. \ref{sch}.
Most studies of the intergalactic medium (IGM) focus on the part of
the spectrum where the only kind of absorption (aside from
occasional absorption by metal systems) is that by
neutral hydrogen via Lyman-alpha (Ly$\alpha$) .i.e. 
the well-known Ly$\alpha$ forest. This is the part of the spectrum
that has a wavelength between $(1+z_q) \lambda_\beta^0$ and $(1+z_q) \lambda_\alpha^0$, where
$z_q$ is the quasar redshift. (Hereafter, we will use the symbols
$\lambda_\alpha^0 = 1215.67 \AA$, $\lambda_\beta^0 = 1025.72 \AA$, and 
$\lambda_\gamma^0 = 972.54 \AA$ to denote the rest-frame Ly$\alpha$, Ly$\beta$
and Ly$\gamma$ wavelengths.)
Absorption blueward of $(1+z_q) \lambda_\beta^0$ 
is generally ignored because Ly$\alpha$ absorption is entangled with absorption
by the higher Lyman series (for exceptions, see e.g. Press, Rybicki \& Schneider 1993). 
For instance, absorption at a wavelength $\lambda$ between $(1+z_q) \lambda_\gamma^0$ and 
$(1+z_q) \lambda_\beta^0$
has two sources:
\begin{eqnarray}
e^{-\tau_{\rm tot.}(\lambda)} = e^{-\tau_\alpha (z_\alpha)} e^{-\tau_\beta (z_\beta)}
\end{eqnarray}
where $\tau_{\rm tot.}$ is the net total optical depth observed at the wavelength $\lambda$, 
and $\tau_\alpha$ and $\tau_\beta$ are the Ly$\alpha$ and Ly$\beta$ optical depths
respectively, which arise from absorption by materials at two different redshifts:
$z_\alpha = [\lambda/\lambda_\alpha^0] - 1$, and $z_\beta = [\lambda/\lambda_\beta^0] - 1$. 
We will refer to this part of the spectrum as the Ly$\alpha + \beta$ region.

The key observation of this paper is that $z_\alpha$ and $z_\beta$ are associated with physically
widely separated parts of the IGM: 
\begin{equation}
u \sim 2 c \times (\lambda_\alpha^0 - \lambda_\beta^0)/(\lambda_\alpha^0 + 
\lambda_\beta^0)  = 5.1 \times 10^4 {\,\rm km/s}
\label{usep}
\end{equation}
where $c$ is the speed of light.
The velocity separation $u$ is much larger than the correlation scale at redshifts
of a few: $\sim 100$ km/s. This means that to good approximation, the Ly$\alpha$
and Ly$\beta$ optical depths that contribute to the total $\tau_{\rm tot.}$ at a
given observed wavelength are uncorrelated. This implies
\begin{eqnarray}
\label{factorize}
&& \langle e^{-\tau_{\rm tot.} (\lambda)} \rangle = \langle e^{-\tau_\alpha (z_\alpha)} \rangle
\langle e^{-\beta_\alpha (z_\beta)} \rangle \, , \\ \nonumber
&& \langle e^{-\tau_{\rm tot.} (\lambda^1)} e^{-\tau_{\rm tot.} (\lambda^2)} \rangle
\\ \nonumber 
&& = \langle e^{-\tau_\alpha (z_\alpha^1)} e^{-\tau_\alpha (z_\alpha^2)} \rangle
\langle e^{-\tau_\beta (z_\beta^1)} e^{-\tau_\beta (z_\beta^2)} \rangle \, ,
\end{eqnarray}
where $\langle ... \rangle$ denotes ensemble averaging (or, operationally, averaging over
lines of sight), and $\lambda^1$ and $\lambda^2$ refers to two different wavelengths
that reside in the Ly$\alpha + \beta$ region. The
redshifts $z_\alpha^1$ and $z_\alpha^2$ are the corresponding Ly$\alpha$ redshifts,
and similarly, $z_\beta^1$ and $z_\beta^2$ are the corresponding Ly$\beta$ redshifts. 

The first equality in eq. (\ref{factorize}), which states that the average transmission
in the Ly$\alpha+\beta$ region factorizes into two parts, is implicitly 
assumed in the existing work that makes use of Ly$\beta$ absorption (e.g. Cen \& McDonald 2002, Fan et al. 2002, 
Lidz et al. 2002).

The second equality goes one step further: it tells us that the two-point correlation
in the same region also factorizes into two parts: the Ly$\alpha$ correlation and the
Ly$\beta$ correlation respectively. Since both the two-point correlation in
$e^{-\tau_{\rm tot.}}$ and the two-point correlation in $e^{-\tau_\alpha}$ can be measured directly
(the latter from separate lines of sight to quasars at lower redshifts), eq. (\ref{factorize})
tells us that we have a handle on the two-point correlation in $e^{-\tau_\beta}$ as well. 
\footnote{Obviously, the same logic applies to the three-point correlation and
so on, which 
we will not discuss here.}

The crucial point here is {\it not} to separate Ly$\alpha$ and Ly$\beta$ absorption
on an absorption-line by absorption-line basis, which is a challenging task. 
Rather, the strategy is to exploit the property of
uncorrelated absorption to statistically separate out the two sources of absorption
in the Ly$\alpha+\beta$ region of quasar spectra.

Why is the Ly$\beta$ absorption interesting? 
Its utility lies in the smallness of the Ly$\beta$ absorption cross-section:
it is lower by a factor of $5.27$ than the Ly$\alpha$ cross-section.
This means that for a given neutral hydrogen density, the associated Ly$\beta$ optical depth
is a factor of $5.27$ lower than the Ly$\alpha$ optical depth.
Therefore, Ly$\beta$ absorption is more sensitive to structure at higher overdensities compared to Ly$\alpha$
(because Ly$\alpha$ goes saturated before Ly$\beta$). 
As we explain, this allows a better
measurement of the equation of state of the IGM. 
Furthermore, by the same token,  Ly$\beta$ absorption also
offers a better hope for constraining the large scale structure at high redshifts 
($z \gsim 5$) where Ly$\alpha$ absorption is often saturated.

The organization of this paper is as follows. 
In \S \ref{eos}, we demonstrate how the Ly$\beta$ forest is more sensitive
to high overdensities than the Ly$\alpha$ forest, especially 
as far as the equation of state is concerned. 
In \S \ref{method}, we develop the method of statistically separating Ly$\beta$ from Ly$\alpha$ absorption
by rewriting eq. (\ref{factorize}) in Fourier space. 
We demonstrate using a concrete example how the Ly$\beta$ transmission power spectrum differentiates
between different equations of state that have very similar Ly$\alpha$ transmission power spectra.
We end in \S \ref{discuss} with a discussion of {\bf 1.} how a better measurement of the equation of state
might allow stronger constraints on the slope of the mass power spectrum, {\bf 2.} how 
the Ly$\beta$ transmission power spectrum provides a more precise test of the gravitational instability
model of the forest, and yields more stringent constraints on feedback processes, and {\bf 3.} 
corrections to eq. (\ref{factorize}).

\section{The Equation of State of the IGM: \\Lyman-$\alpha$ versus Lyman-$\beta$}
\label{eos}

The photoionized IGM is well described by a temperature-density relation, or
an effective equation of state, of the form:
\begin{equation}
T = T_0 \Delta^\alpha
\label{eosT}
\end{equation}
where $T$ is the temperature, $T_0$ is its value at mean density, $\alpha$ is the slope
of the equation of state, and $\Delta = \rho/\bar\rho$, with
$\rho$ being the gas density, and $\bar\rho$ its mean. Such a relation appears to hold
for $\Delta \lsim 5$ (Miralda-Escude \& Rees 1994, Hui \& Gnedin 1997). 

Current measurements from the Ly$\alpha$ forest in the
redshift range of $z \sim 2.4-4.0$ yield much better constraints on
$T_0$ than on $\alpha$ (Ricotti, Gnedin \& Shull 2000, Schaye et
al. 1999, Bryan \& Machacek 2000, McDonald et al. 2001,
Meiksin, Bryan \& Machacek 2001, Zaldarriaga, Hui \& Tegmark 2001 [ZHT01 hereafter]). 
The reason is quite simple to understand. Ly$\alpha$
absorption is sensitive largely to $\Delta \sim 1 - 2$, and the short lever arm does
not allow a precise measurement of the slope $\alpha$. Present constraints are consistent with
the full physically plausible range $\alpha = 0 - 0.6$ (Hui \& Gnedin 1997), 
according to ZHT01. \footnote{ZHT01 constrain $\alpha$ using measurements
of the small scale flux power spectrum. McDonald et al. (2001), 
using a line--fitting method, find a tighter constraint 
on $\alpha$ at $z=2.4$, requiring
$\alpha \geq 0.38$ at a $1\sigma$--confidence level. Schaye et
al. (2000) obtain still tighter constraints also using a line--fitting
method. The difference between the Schaye et al. (1999) error
bars and the McDonald et al. (2001) error bars is due to different
line selection criteria (See McDonald et al. (2001) for a discussion of this).}

\begin{figure*}[t]
\vbox{ \centerline{ \epsfig{file=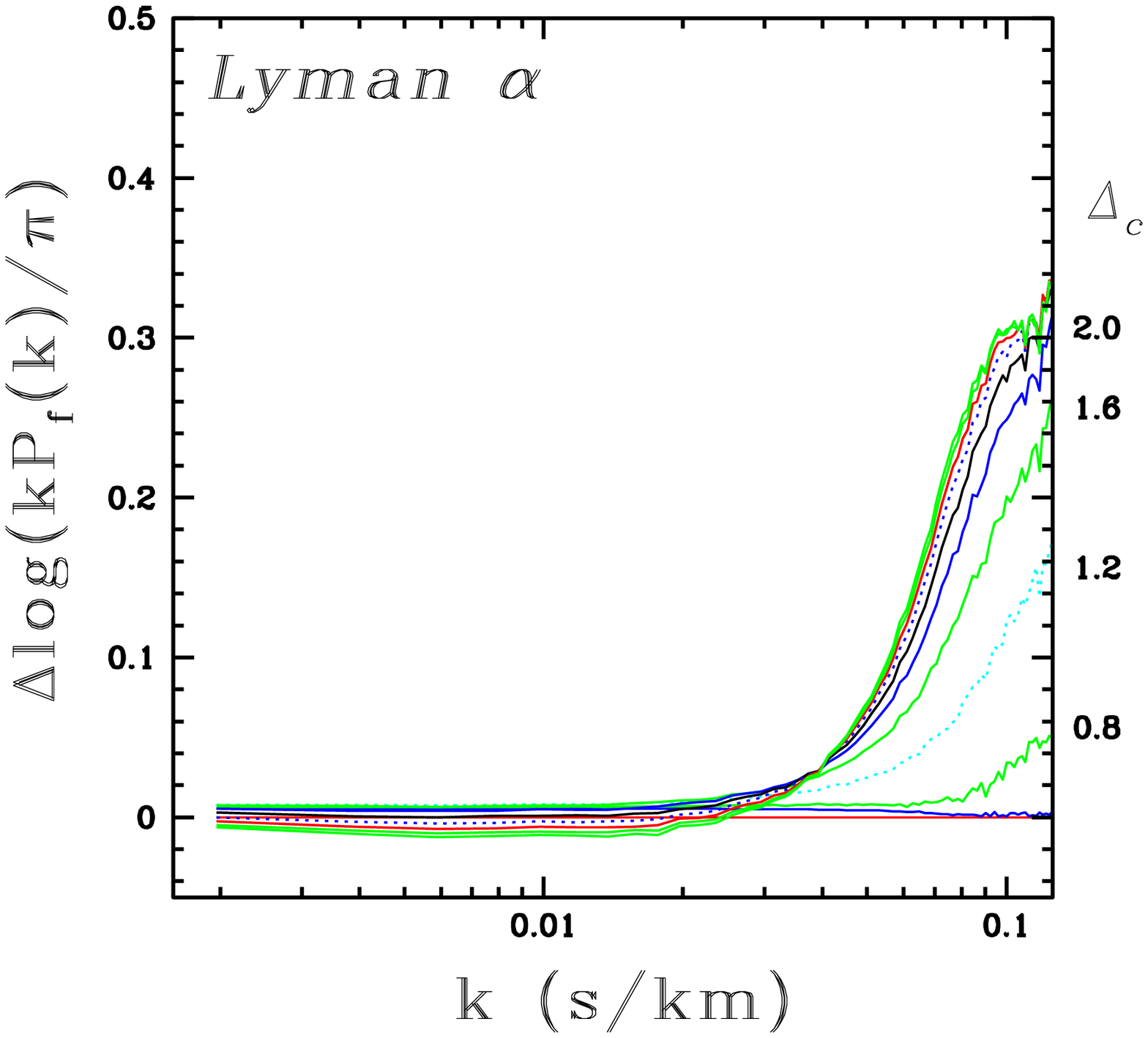,width=9.5truecm}
\epsfig{file=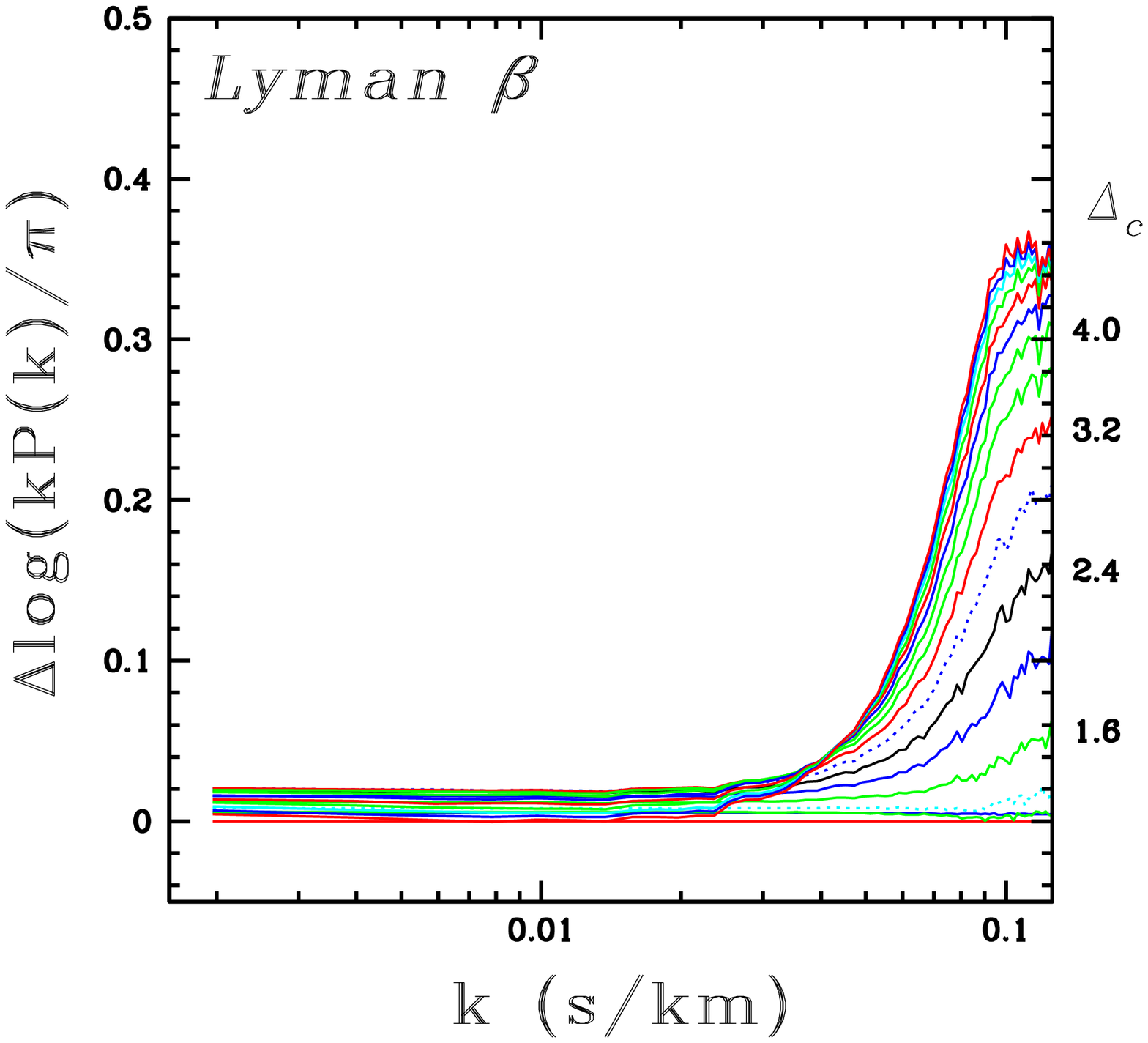,width=9.5truecm}}
\figcaption[]{ \footnotesize Illustration of how Ly$\alpha$
and Ly$\beta$ absorption probes different overdensities.
{\it Left:} Differential Ly$\alpha$ transmission power spectra for a 
set of artificial temperature-density relations defined by eq.
(\ref{eosT}) for $\Delta_c$ ranging from $0$ to $3.6$ in steps of $0.4$ (bottom
to top, as seen around $k = 0.1$ s/km). What is shown in each
case is not the absolute power spectrum, but rather the
logarithmic difference from a baseline power spectrum defined
by that of $\Delta_c = 0$ (this is why the line corresponding
to $\Delta_c = 0$ is exactly zero). 
{\it Right:} Differential Ly$\beta$ transmission power spectra
for $\Delta_c$ ranging from $0$ to $6.0$ in steps of $0.4$.
The redshift in both panels is $z = 3.0$.
 [{\it See the electronic version of the journal for a color version of this figure.}]
\label{deg}}}
\end{figure*}

To illustrate the range of overdensities to which the Ly$\alpha$ forest is
sensitive, we measure the transmission power spectrum from 
mock Ly$\alpha$ spectra that are generated using N-body simulations.
The simulations used throughout this paper are $256^3$-grid, 
$128^3$-particle, P3M simulations of a Standard Cold-Dark-Matter model 
with a box size of $16$ Mpc/h. The simulations are dark-matter only
simulations generated with the HYDRA code of Couchman et al. (1995).
The baryon density is obtained by smoothing the dark matter density to mimic
the effect of pressure forces (See ZHT01 for more details). These 
simulations have lower resolution than recommended by McDonald (2003) and 
Meiksin \& White (2003) to achieve
convergence in measuring the flux power spectrum. We expect, however, 
that our present simulations are adequate to qualitatively 
illustrate the benefits of measuring the Ly$\beta$ flux power spectrum.

In Figure 2 ({\it left panel}) we show the Ly$\alpha$ transmission power spectrum at $z=3.0$ for a set
of artificial equations of state parameterized in the form (following ZHT01):
\begin{equation}
T=\left\{\begin{array}{ll} T_0 \hspace{5mm} \rm{for} \hspace{2mm} \Delta < \Delta_c\\
 2T_0  \hspace{3.3mm} \rm{for} \hspace{2mm} \Delta > \Delta_c
\end{array}\right.
\end{equation}

where $\Delta_c$ defines the density threshold where the gas temperature $T$
has a jump and $T_0=1.2 \times 10^4$ K. The left panel shows the Ly$\alpha$ transmission power spectrum
for $\Delta_c$ varying from $0.0$ to $3.2$ in steps of $0.4$. 
We have subtracted from each power spectrum a baseline power spectrum 
corresponding to that of $\Delta_c = 0.0$. In comparing power spectra
with different $\Delta_c$, we fix all of the other IGM modeling parameters.

The Ly$\alpha$ transmission (or flux) power spectrum $P_{f,\alpha} (k)$ is defined as:
\begin{eqnarray}
\label{Pkalpha}
\xi_{f,\alpha} (u) = \langle \delta_{f,\alpha} (u^1) \delta_{f,\alpha} (u^1 + u) \rangle \, , \\ \nonumber
P_{f,\alpha} (k) = \int {dk\over 2\pi} \xi_{f,\alpha} (u) e^{-iku}
\end{eqnarray}
where $\xi_{f,\alpha} (u)$ is the two-point correlation function at a velocity
separation of $u$, the transmission power spectrum $P_{f,\alpha} (k)$ is its
Fourier counterpart as a function of wavenumber $k$, 
and $\delta_{f,\alpha}$ is the fluctuation of the Ly$\alpha$ transmission
about its mean:
\begin{eqnarray}
\label{deltafalpha}
\delta_{f,\alpha} \equiv {e^{-\tau_\alpha} \over \langle e^{-\tau_\alpha} \rangle} - 1 \, .
\end{eqnarray}
Note that the definition of $\xi_{f,\alpha}$ differs from the kind of two-point correlation
in eq. (\ref{factorize}) in that $\xi_{f,\alpha}$ involves the correlation of 
$\delta_{f,\alpha}$, which has zero mean, and is properly normalized.
We will sometimes refer to $\xi_f$ as the normalized two-point function.

Figure. 2 ({\it left panel}) shows that the Ly$\alpha$ transmission power spectrum 
at high $k$ ($\sim 0.1$ s/km) increases, as $\Delta_c$ increases from $0$ (or, in other words,
according to eq. (\ref{eosT}), as the temperature drops). This is consistent with
the picture that thermal broadening smooths the flux field, and suppresses small scale power, a fact that
is used to constrain the thermal state of the IGM from observations (ZHT01). 
The interesting point of Figure 2 ({\it left panel}) is that
changes in the Ly$\alpha$ transmission power spectrum are most pronounced in a narrow range of
$\Delta_c$'s, from about $1$ to $2$. For smaller or larger $\Delta_c$'s, the 
Ly$\alpha$ power spectrum does not change in a substantial way. 
This is fundamentally why current observations from the Ly$\alpha$ forest 
yield useful constraints on the temperature around mean density (or $1-2$ times the mean
density), but not on the slope of the equation of state $\alpha$ (eq. (\ref{eosT}) ). 
A stronger constraint can be obtained if one has the means to probe the temperature
of the IGM over a larger range of densities.

The Ly$\beta$ transmission power spectrum provides the requisite probe of the
IGM over a large range of densities, 
as illustrated by Figure 2 ({\it right panel}). This panel is similar to
the left panel, except that it shows the Ly$\beta$ transmission power spectrum
rather than the Ly$\alpha$ power spectrum.  The Ly$\beta$ transmission power is
defined by (similar to eq. (\ref{Pkalpha}), and
(\ref{deltafalpha}) ):
\begin{eqnarray}
\label{Pkbeta}
\xi_{f,\beta} (u) = \langle \delta_{f,\beta} (u^1) \delta_{f,\beta} (u^1 + u) \rangle \, , \\ \nonumber
P_{f,\beta} (k) = \int {dk\over 2\pi} \xi_\beta^f (u) e^{-iku} \, ,
\end{eqnarray}
\begin{eqnarray}
\label{deltafbeta}
\delta_{f,\beta} \equiv {e^{-\tau_\beta} \over \langle e^{-\tau_\beta} \rangle} - 1 \, .
\end{eqnarray}
In the right panel, we let $\Delta_c$ vary from $0.0$ at the bottom to $6.0$ at the top.
Clearly, the Ly$\beta$ transmission power spectrum is sensitive to higher densities: 
it continues to vary from $\Delta_c$ of $1$ to about $4$. 
\footnote{At sufficiently high overdensities, the equation of state given in
eq. (\ref{eosT}) would no longer be a good description of the temperature-density
relation, because large--scale shock heating becomes important, which introduces a significant
scatter to the temperature. The temperature-density relation for
overdensities of $\Delta$ up to $\sim 5$ can probably still be described
by eq. (\ref{eosT}) (Hui \& Gnedin 1997).
}

Note that the experiment shown in Fig. 2. is done at $z = 3$. 
If one moves to higher redshifts, the $\Delta$ one is sensitive to
shift to lower values for both Ly$\alpha$ and Ly$\beta$. This is simply
because the mean density of the universe is higher, and lower overdensities
give rise to the same amount of absorption as at $z=3$. 
For instance, we have repeated the above experiment at $z = 3.74$, and
find that the range of sensitive $\Delta$ shifts downward by about $0.5$
for both Ly$\alpha$ and Ly$\beta$. 

Fig. 2. is only meant to illustrate the range of densities
to which Ly$\alpha$ and Ly$\beta$ absorption are sensitive. 
Does the Ly$\beta$ transmission power spectrum differentiate between realistic 
equations of state that are otherwise difficult to disentangle using the
Ly$\alpha$ power spectrum alone? 
This is addressed in Fig. 3.
At the top of the figure, we show three different Ly$\alpha$ transmission power spectra,
 each assuming a different equation of state slope ($\alpha$ in eq. (\ref{eosT}) ).
The other IGM model parameters (see ZHT01) are adjusted slightly 
(well within observational uncertainties) 
\footnote{In the notation of ZHT01, model A/B  has 
$(k_F,\alpha,T_0,\bar{f})=(36, 0.0, 310, 0.512)$ / $(39, 0.4, 305, 0.507)$.
Here $k_F$ is in units of $h$ Mpc$^{-1}$ and $T_0$ is in units of $(km/s)^2$.}
in each case to give very similar $P_{f,\alpha} (k)$'s. 
The Ly$\beta$ transmission power spectra for these three cases show more differences
at high $k$, suggesting that observational constraints on $P_{f,\beta} (k)$ might
be useful. Our next task is to describe how to tease out $P_{f,\beta} (k)$ from
the Ly$\alpha+\beta$ region of quasar spectra.

\begin{figure}[t]
\vbox{ \centerline{ \epsfig{file=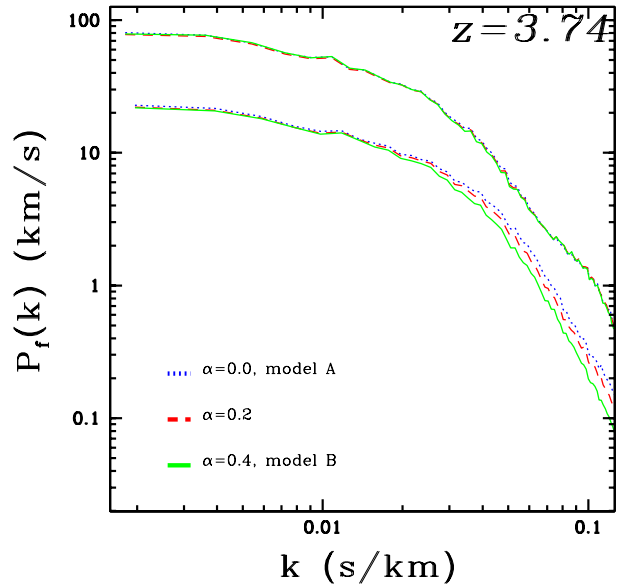,width=10truecm}}
\figcaption[]{ \footnotesize Transmission power spectra for Ly$\alpha$ ({\it top three curves})
and Ly$\beta$ ({\it bottom three curves}) for three models that have different equations of state.
The models are chosen so that the Ly$\alpha$ power spectra are very similar.
This is at $z = 3.74$.[{\it See the electronic version of the Journal for a color version of this figure}]
\label{fig3deg}}}
\end{figure}

\section{Constraining the Ly$\beta$ Power Spectrum}
\label{method}

As discussed in \S \ref{intro}, the coincident Ly$\alpha$ and Ly$\beta$ 
absorption in the Ly$\alpha+\beta$ region of a quasar spectrum are to a good approximation uncorrelated. This gives rise to a simple factorization of the (un-normalized) 
two-point function (eq. (\ref{factorize}) ).
The normalized two-point function obeys
\begin{eqnarray}
\label{xifactorize}
\xi_{f,{\rm tot.}} (u) = \xi_{f, \alpha} (u) + \xi_{f, \beta} (u) + 
\xi_{f, \alpha} (u) \xi_{f, \beta} (u)
\end{eqnarray}
where $\xi_{f,{\rm tot.}}$ as a function of velocity separation $u$ is defined in a 
similarly to eq. (\ref{Pkalpha}) and (\ref{Pkbeta}) 
(i.e.  $\xi_{f,{\rm tot.}} (u) \equiv \langle \delta_{f,{\rm tot.}} (u_1) \delta_{f,{\rm tot.}} (u_1 + u) \rangle$,
with $\delta_{f,{\rm tot.}} \equiv [e^{-\tau_{\rm tot.}} /\langle e^{-\tau_{\rm tot.}} \rangle] -1$, and
$\tau_{\rm tot.}$ being the total Ly$\alpha$ + Ly$\beta$ optical depth).
The velocity separation $u$ is related to the
quantities $\lambda^1$, $\lambda^2$, $z_\alpha^1$ $z_\alpha^2$, $z_\beta^1$ and $z_\beta^2$ in eq. (\ref{factorize}) by:
\begin{eqnarray}
u = c {\lambda^2 - \lambda^1 \over \bar\lambda} = 
c {z_\alpha^1 - z_\alpha^2 \over 1 + \bar z_\alpha} =
c {z_\beta^1 - z_\beta^2 \over 1 + \bar z_\beta}
\end{eqnarray}
where $\bar\lambda$ is the mean observed wavelength in the Ly$\alpha+\beta$ region,
and $\bar\lambda = (1+\bar z_\alpha) \lambda_\alpha^0 = (1+\bar z_\beta) \lambda_\beta^0$.

The Fourier counterpart of eq. (\ref{xifactorize}) is
\begin{eqnarray}
\label{Pkfactorize}
P_{f,{\rm tot.}} (k) &=& P_{f,\alpha} (k) + P_{f,\beta} (k) \\ \nonumber 
&+& \int {dk'\over 2\pi} P_{f,\alpha} (k-k') P_{f,\beta} (k')
\end{eqnarray}
where $P_{f,{\rm tot.}}$ is the power spectrum
of the total transmission in the Ly$\alpha+\beta$ region. We will often refer to $P_{f,{\rm tot.}}$ as the total power.
Note that implicit in the above expression is that
$P_{f,\alpha}$ and $P_{f,\beta}$ are at different mean redshifts: $\bar z_{\alpha}$ and 
$\bar z_{\beta}$. 

One can directly measure both $P_{f, {\rm tot.}}$ and $P_{f,\alpha}$ from 
observational data. To constrain the Ly$\beta$ power spectrum $P_{f,\beta}$, it
is important that quasars at different redshifts are employed to measure 
$P_{f, {\rm tot.}}$ and $P_{f,\alpha}$ -- the former from higher redshift quasars, 
and the latter from lower redshift quasars whose
Ly$\alpha$-only region overlaps in wavelengths with the Ly$\alpha+\beta$ region of the former.

In principle, once $P_{f, {\rm tot.}}$ and $P_{f,\alpha}$ are given, 
eq. (\ref{Pkfactorize}) can be inverted to obtain $P_{f,\beta}$ if one
thinks of it as a linear vector equation:
\begin{eqnarray}
\label{PM}
{\bf P_{f, {\rm tot.}}} - {\bf P_{f, \alpha}} = {\bf M} \cdot {\bf P_{f, \beta}}
\end{eqnarray}
where ${\bf M}$ is a matrix whose components are
\begin{eqnarray}
M(k_i, k_j) = \delta_{ij} + {dk_j \over 2\pi} P_{f,\alpha} (k_i - k_j) \, .
\end{eqnarray}

In other words, from eq. (\ref{PM}), one can in principle obtain:
\begin{eqnarray}
{\bf P_{f, \beta}} = {\bf M}^{-1} \cdot ({\bf P_{f, {\rm tot.}}} - {\bf P_{f, \alpha}})
\end{eqnarray}
While such an inversion is useful for visually inspecting the Ly$\beta$
power spectrum, in practice it can be noisy and one is likely better
off focusing on the total power but keeping in mind that the Ly$\alpha$
contribution is known. 

We therefore will not pursue the path of inversion here. Instead, we will be content
with posing the question: how different is the total observable
power $P_{f,{\rm tot.}}$ for models that are quite degenerate
in their Ly$\alpha$ power, for instance those shown in Fig. 3?

\begin{figure}[t]
\vbox{ \centerline{ \epsfig{file=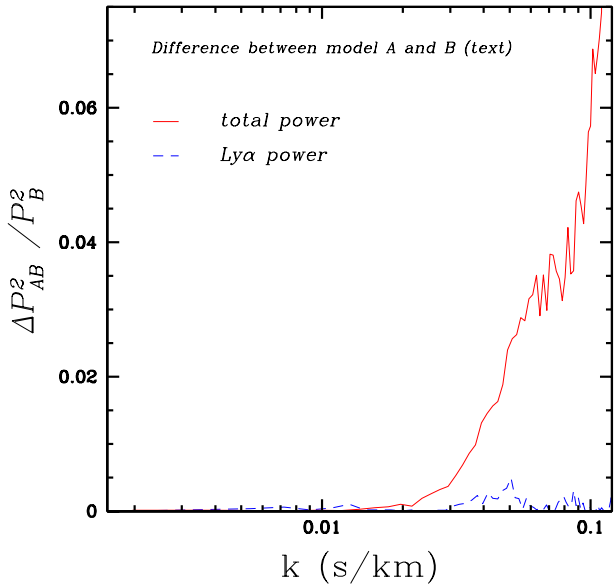,width=10truecm}}
\figcaption[] 
{ \footnotesize Fractional differences (squared) between the two models 
A and B (with the slope of the equation of state $\alpha = 0.0$ and $\alpha = 0.4$ respectively;
see Fig. 2.) in the total and Ly$\alpha$ power spectra.
[{\it See the electronic version of the Journal for a color version of this figure}]
\label{fig:tot}}}
\end{figure}

The answer is provided by Fig. \ref{fig:tot}. 
Shown here is the fractional difference (squared) in the total 
power ({\it top curve}) between
the two models labeled $A$ and $B$ ($\alpha = 0$ and $\alpha = 0.4$ respectively) in Fig. 3.
The total power here is computed using eq. (\ref{Pkfactorize}) by taking
$P_{f,\beta}$ at $\bar z_\beta = 3.74$, and $P_{f,\alpha}$ at $\bar z_\alpha = 3.0$.

For comparison, we show in the same figure the fractional difference in the
Ly$\alpha$ power ({\it bottom curve}) between the same two models at the same redshift as
the Ly$\beta$ power, i.e. $z = 3.74$. Comparing the two curves tells us
how much more the total power (due to the information from
Ly$\beta$ absorption) can constrain the equation of state at $z = 3.74$ 
compared
to the Ly$\alpha$ power spectrum alone. We assume implicitly here 
that $P_{f,\alpha}$ at $\bar z_\alpha = 3.0$ is well known from data at low redshifts.
\footnote{In order to calculate the total power, we use the same 
$P_{f, \alpha}$ at $\bar z_\alpha = 3.0$ in model A and model B 
-- the difference in the total power of these models comes only
from differences in their Ly$\beta$ power.}

From Fig. \ref{fig:tot}, we can infer the signal-to-noise ratio (S/N) with
which one can distinguish between the two different equations
of state (models A and B). 
The $({\,\rm S/N})^2$ is given by
\begin{eqnarray}
\label{SN}
(S/N)^2 = \sum_k \big{[}P^A_f (k) - P^B_f (k)\big{]}^2\Big{/}\hspace{1mm} \sigma_P (k)^2
\end{eqnarray}
where $P^A_f$ and $P^B_f$ are the power spectra for model 
$A$ and $B$ respectively ($P_f$ here can stand for
either $P_{f,\alpha}$ or $P_{f,{\rm tot.}}$), and $\sigma_P (k)^2$ is the variance
in power (here taken to be that for model $B$). The summation
extends to all modes with positive $k$ values, up to $k_{\rm max}=0.12$ s/km. 
The power at higher $k$ is expected to be contaminated significantly by
metal lines (McDonald et al. 2000) and so the sum is truncated at 
$k_{\rm max}$. The variance $\sigma_P (k)^2$ equals
$(P^B_f (k) + {\,\rm shot})^2$, if
one approximates the fluctuations as Gaussian random (see Hui et al. 2001).
It is likely inaccurate to assume Gaussian errors for some of the high 
$k$ modes considered here. A more accurate error estimate is, however, unlikely
to change our main point that the Ly$\beta$ flux power spectrum distinguishes
between models with different temperature-density relations, $\alpha$, that
have identical Ly$\alpha$ flux power spectra.

The shot-noise term 'shot' is approximately equal to 
$(\Delta u/\langle e^{-\tau}\rangle) (s/n)^{-2}$, where $\Delta u$ is
the size of a pixel in km/s, and $s/n$ is the signal-to-noise ratio per pixel of
the data (different from $S/N$ defined above). 
High quality spectra (e.g. Kim et al. 1997, Kirkman \& Tytler 1997, 
Simcoe, Sargent \& Rauch 2002, Kim et al. 2002) often have a sufficiently
low shot-noise that shot-noise can be ignored altogether, in which
case, $(S/N)^2 = \sum_k (P^A_f (k) - P^B_f (k))^2/ P^B_f (k)^2$, which is the area under the curve shown in Fig. (\ref{fig:tot}).

We find that using the Ly$\alpha$ power spectrum at a mean redshift of
$z = 3.74$ gives a discriminating power of $(S/N)^2 = 0.8$, while
using the total power from the Ly$\alpha+\beta$ region 
(corresponding to Ly$\beta$ at the same $z = 3.74$) gives $(S/N)^2 = 7.6$.
This is the $(S/N)^2$ for one line of sight, assuming that 
the Ly$\alpha$ and Ly$\alpha+\beta$ regions have a length of $5.1
 \times 10^4$ and $ 5.1 \times 10^4(\lambda_\alpha^0-\lambda_\beta^0)/
(\lambda_\beta^0-\lambda_\gamma^0)=1.4 \times 10^4$ km/s
respectively. The discriminating power $(S/N)^2$ scales linearly with the 
number of line of sights used. Note that this estimate assumes that one 
has enough resolution to
measure the power spectra at high $k$ (up to $k = 0.12$ s/km) 
i.e. a resolution of Full-Width-at-Half-Maximum (FWHM) $= 8$ km/s or better.
In summary, using the Ly$\beta$ forest boosts the discriminating power
($S/N$) between an equation of state of $\alpha = 0$ and an equation of
state $\alpha = 0.4$ by a factor of $\sim 3$. If only modes up to 
$k = 0.1$ s/km are included then this number goes down to $2.5$. These numbers
are typical for models very close in Ly$\alpha$.

To completely {\it quantify} how much better one can constrain the equation of 
state, $\alpha$, using the Ly$\beta$ flux power spectrum in addition
to the Ly$\alpha$ flux power spectrum, a more rigorous approach
is necessary. One should run a large grid of simulated models, simulating
both $P_{f,{\rm tot.}}$ and $P_{f,\alpha}$, compare with mock data, and
marginalize over all of the other modeling parameters to obtain
the reduced likelihood function for $\alpha$. A comparison between the
resulting likelihood function formed using $P_{f,\alpha}$ alone, with
that formed using $P_{f,{\rm tot.}}$ plus $P_{f,\alpha}$, quantifies how much one gains
using the Ly$\beta$ forest. Here we are content to illustrate that the
Ly$\beta$ flux power spectrum adds extra information on $\alpha$ that is not
available from the Ly$\alpha$ flux power spectrum alone (Figure 3), 
without completely quantifying how much tighter the resulting constraints on $\alpha$ are.

\section{Discussion}
\label{discuss}

The discussion \S3 suggests that the Ly$\beta$ forest
can indeed be beneficially used in conjunction with the Ly$\alpha$ forest to
help discriminate between different models, especially between
different equations of state. The reason for its utility lies
in its sensitivity to higher overdensities. 
Better measurements of the equation of state are useful in constraining
the reionization history of the universe (Theuns et al. 2002, Hui \& Haiman 2003).
Since spectra of quasars at sufficiently high redshifts 
often extend well into the Ly$\beta$ forest, there is no reason not to exploit this
part of the spectrum to increase the scientific return.
An alternative possibility for constraining the equation of state is
to use the Ly$\alpha$ flux bispectrum in conjunction with the flux
power spectrum. Referring to Table 3 of Mandelbaum et al. (2003),
it appears that combining the flux bispectrum and power spectrum
yields tighter constraints on the temperature density
relation than using the power spectrum alone. 

Zaldarriaga, Scoccimarro \& Hui (2001) pointed out that from the Ly$\alpha$ power spectrum
alone, there is a near-degeneracy between models that trade off variations in the
mass power spectral index $n_s$ with variations in the equation of state.
Our finding that the Ly$\beta$ forest can place stronger constraints
on the equation of state suggests that one might be able to break this degeneracy 
using the Ly$\beta$ power spectrum.

We have carried out an experiment similar to that in Fig. \ref{fig:tot}, except we
replace models A and B with the following two models: one has $n_s = 0.7$, $\alpha = 0.0$ and
$T_0 = 2.1 \times 10^4$ K, and the other has $n_s = 1.1$, $\alpha = 0.6$ and 
$T_0 = 1.6 \times 10^4$ K
(see eq. (\ref{eosT}) for definitions of $\alpha$ and $T_0$);
these two models have quite similar Ly$\alpha$ power spectra. 
The total power distinguishes between the models at a level similar to that shown in Fig. 4.
However a perhaps more readily realizable option is to break the degeneracy by measuring 
the Ly$\alpha$ power spectrum as accurately as possible on large scales, $k \sim 0.001 - 0.01$ s/km. 
In this range the models {\it are} different in the Ly$\alpha$ power spectrum, albeit at a level
that is too small to distinguish with existing data.
Quasar spectra from the Sloan Digital Sky Survey are well suited for this.

An interesting use of the Ly$\beta$ forest is to search for signs of feedback
processes in the IGM. Recent interest in the Ly$\alpha$ forest as a
cosmological probe relies on a framework in which fluctuations in 
the forest arise naturally from gravitational instability 
(e.g. Bi, Borner, \& Chu 1992, 
Cen et al. 1994, Zhang et al. 1995, Reisenegger \& Miralda-Escude 1995, Hernquist et al. 1996, 
Miralda-Escude et al. 1996, Muecket et al. 1996, Bi \& Davidsen 1997, Bond \& Wadsley 1997,
Rauch et al. 1997, Hui, Gnedin \& Zhang 1997, Croft et al. 1998, Theuns et al. 1999, 
Nusser \& Haehnelt 2000,
McDonald et al. 2000, White \& Croft 2000, Meiksin et al. 2001, Pichon et al. 2001, 
Croft et al. 2002, Gnedin \& Hamilton 2002, Viel et al. 2002).
An important assumption behind this picture is that feedback processes, such
as galactic winds (Adelberger et al. 2003), do not significantly disturb the IGM. 
The good agreement between observations (particularly the Ly$\alpha$ transmission power spectrum)
and the gravitational instability model is often used as an argument that 
feedback processes, while inevitably present, do not affect large volumes of the IGM.
A reasonable expectation is that they preferentially affect higher density regions.
If so, the Ly$\beta$ forest offers a better hope of testing for the presence
of such feedback processes. Particularly interesting is the fact that once the 
gravitational instability model parameters
(such as the mass power spectrum, cosmology, etc) are tuned to match observations
of the Ly$\alpha$ forest, there are definite predictions for the correlations observed
in the Ly$\beta$ forest, since Ly$\alpha$ and Ly$\beta$ optical depths are simply
related by a rescaling in the cross-section. 
As we have discussed, changing the slope of the equation of state ($\alpha$)
does seem to modify the Ly$\beta$ power spectrum while leaving the Ly$\alpha$
power spectrum relatively unchanged. Feedback 
processes might behave in the same way i.e. affecting the Ly$\beta$ forest more
than the Ly$\alpha$ forest, except that it is unlikely that their effects can be mimicked by simply
varying $\alpha$. Galactic winds for instance change the density structure of
the IGM, by creating evacuated shells around galaxies. They might also introduce a larger than expected
scatter in the IGM temperature at high overdensities. 

It is important to reiterate the method we advocate is {\it not} to 
decipher the Ly$\beta$ forest on an absorption line--by--absorption-line basis.
Rather, the strategy is to statistically detect the presence of Ly$\beta$ correlation,
exploiting the fact that the Ly$\alpha$ and Ly$\beta$ absorptions that fall within the Ly$\alpha+\beta$ region
of a quasar spectrum are uncorrelated (eq. (\ref{factorize}) ).
A natural question is: how good an approximation is it?
The fractional correction to the first equality of eq. (\ref{factorize})
is the two-point correlation between Ly$\alpha$ and Ly$\beta$
transmission fluctuations at a velocity separation of 
$u \sim 5.1 \times 10^4$ km/s
(eq. (\ref{usep}) ). It is safe to assume that this correlation is
weaker than the Ly$\alpha$ two-point correlation ($\xi_{f,\alpha}$; see
eq. (\ref{Pkalpha}) ), since we know from experience that Ly$\beta$ absorption
weakens the correlation (compare the Ly$\alpha$ and Ly$\beta$ curves in
Fig. 3). The observed two-point correlation function in the Ly$\alpha$ forest
has only been reliably measured at velocity separations of 
$u \lesssim 1700$ km/s (McDonald et al. 2000). 
In order to estimate the two-point correlation
function at velocity separations of $u \sim 5.1 \times 10^4$ km/s, we
extrapolate from our simulation measurements assuming linear biasing
(Scherrer \& Weinberg 1998, McDonald et al. 2000, Lidz et al. 2002).
The resulting estimate is 
$\xi_{f,\alpha} (u=5.1 \times 10^4 {\,\rm km/s}) \lesssim 10^{-5}$.
Therefore, any correction to the first part of 
equation (\ref{factorize}),
$\langle e^{-\tau_{\rm tot.} (\lambda)} \rangle = \langle e^{-\tau_\alpha (z_\alpha)} \rangle
\langle e^{-\beta_\alpha (z_\beta)} \rangle$, must be very small: 
$\lesssim 10^{-5}$. 
Such a correction is even smaller at lower redshifts where the two-point correlation
is weaker. We can similarly estimate the corrections to the factorization
of the two-point function (second part of eq. (\ref{factorize}), or
equivalently, eq. (\ref{xifactorize}) ). The fractional error we make in the
two point function is roughly
$\sim 2  \xi_{f, \alpha \beta} (u_{\rm \alpha \beta}, \bar z_{\rm \alpha
\beta}) / \xi_{f, \beta} (u, \bar z_{\beta})$. Here 
$\xi_{f, \alpha \beta}$ refers to the two point correlation between  
the Ly$\beta$ absorber that absorbs at observed wavelength 
$\lambda^2$ and the Ly$\alpha$ absorber 
that absorbs at $\lambda^1$, $\bar z_{\rm \alpha \beta}$ is the mean redshift 
between these absorbers, and $u_{\rm \alpha \beta}$ is their velocity 
separation. 
The fractional error depends on the relative size of $u$ and
$u_{\rm \alpha \beta}$. Here we estimate the error when the Ly$\beta$
absorbers are separated by one correlation length, $u \sim 100$ km/s, and
situated at $\bar z_{\beta}=3.74$.  In this case the Ly$\alpha$ absorber
is separated from the Ly$\beta$ absorber by 
$u_{\rm \alpha \beta} \sim 5.1 \times 10^4$ km/s. A conservative error
estimate then comes from taking $\xi_{f, \beta} (u, \bar z_{\beta}) 
\sim \xi_{f, \alpha} (u, \bar z_{\beta})/5$, (see Fig. 3),and 
$\xi_{f, \alpha \beta} (u_{\rm \alpha \beta},\bar z_{\rm \alpha \beta}) \sim \xi_{f, \alpha} (u_{\rm \alpha \beta},\bar z_{\beta})/\sqrt{5}$.
From McDonald et al. (2000), the observed two-point function at a separation
of one correlation length is
$\xi_{f, \alpha} (u \sim 100 {\,\rm km/s}, \bar z \sim 4) \sim 0.2$,
which shows that the fractional error is 
$\lesssim (10/\sqrt{5})(10^{-5}/0.2) \sim 0.03\%$.
 For very widely spaced pixels in the quasar spectrum, the fractional error is larger, but
these pixels have a negligible correlation.

Two issues are worth further exploration. First, one must take care in 
masking out the intervening metal absorption lines in the Ly$\alpha+\beta$ 
region of the spectrum, just as one normally does in the Ly$\alpha$ forest.
Metal lines that cannot be easily masked out, such as OVI, can be suitably
divided out using a method similar to the one mentioned in this paper
i.e. use lines of sight where the red-side of Ly$\alpha$ coincides in wavelengths
with the Ly$\beta$ forest of interest; the two-point correlation of metal
absorption from these lines of sight can be used to take out the metal
contamination, much as we remove the Ly$\alpha$ contamination 
to the Ly$\beta$ forest (a similar technique for 'cleaning' the Ly$\alpha$
forest has been developed by P. McDonald and U. Seljak [2003, private communication]).
Second, it is interesting to explore whether the ideas presented here can
be extended to the higher Lyman series. While this is in principle possible, 
it is likely that the combination of diminishing path lengths, and the
increasing entanglement of different kinds of absorption (i.e. the coexistence
of Ly$\alpha$, $\beta$, $\gamma$, and so on), makes it difficult to exploit
the factorization of correlations in practice.

{\bf Acknowledgments} MD and AL thank the Theoretical Astrophysics group
at Fermilab for their hospitality. We thank Arlin Crotts for useful
conversations and for providing the spectrum of Q2139-4434 for figure 1, 
and Matias Zaldarriaga for making his simulations available to us. We also
thank the referee for useful comments, in particular for 
pointing out the importance of OVI and the possibility
of constraining the equation of state using the flux bispectrum in 
conjunction with the flux power spectrum.
This work is supported in part by an Outstanding
Junior Investigator Award from the DOE, by NSF grant AST-0098437
and the NASA grant NAG5-10842.

\end{document}